\documentclass[manuscript]{raa}
\usepackage{graphicx,times}             %for PS/EPS graphics inclusion, new

\begin{document}

   \title{Can the temperature of Ellerman Bombs be more than 10000K?
%\footnotetext{$*$ Supported by the National Natural Science Foundation of China.}
}

   \volnopage{Vol.0 (200x) No.0, 000--000}      %%preserved for Editor. DOn't remove!
   \setcounter{page}{1}          %%starting page, preserved for Editor. DOn't remove!

   \author{C. Fang
      \inst{1,2,3}
   \and Q. Hao
      \inst{1,2,3}
   \and M. D. Ding
      \inst{1,2,3}
   \and Z. Li
      \inst{1,2,3}
   }

   \institute{School of Astronomy and Space Science, Nanjing
       University, Nanjing 210093, China; {\it fangc@nju.edu.cn}\\
        \and
             Key Laboratory of Modern Astronomy and Astrophysics
             (Nanjing University), Ministry of Education, China;\\
        \and
             Collaborative Innovation Center of Modern Astronomy and Space Exploration
   }

   \date{Received~~2017 month day; accepted~~2017~~month day}

\abstract{
  Ellerman bombs (EBs) are small brightening events in the solar lower atmosphere.
By original definition, the main EB's characteristic is the two emission bumps in
both wings of chromospheric lines, such as H$\alpha$ and \ion{Ca}{II} 8542 {\AA} lines.
Up to now, most authors found that the temperature increase of EBs around
the temperature minimum region is in the range of 600K-3000K. However, with recent IRIS
observations, some authors proposed that the temperature increase of EBs could be more
than 10000K. Using non-LTE semi-empirical modeling, we investigate the line profiles,
continuum emission and the radiative losses for the EB models with different temperature increases,
and compare them with observations. Our result indicates that if the EB maximum temperature attains
more than 10000K around the temperature minimum region, then the resulted H$\alpha$ and
\ion{Ca}{II} 8542 {\AA} line profiles and the continuum emission would be much stronger than
that of EB observations. Moreover, due to the high radiative losses, the high temperature EB
would have a very short lifetime, which is not comparable with the observations. Thus, our study
does not support the proposal that the EB temperatures are higher than 10000K.
}


 \authorrunning{C. Fang, Q. Hao, et al.}            %author_head in even pages
   \titlerunning{Temperature of Ellerman Bombs}  % title_head in odd pages

   \maketitle
\section{Introduction}\label{sect1}

Ellerman bombs (EBs: Ellerman~\cite{Ell17}) are small-scale
brightening events in the solar lower atmosphere. Their characteristic
feature is the excess emission in the wings of chromospheric lines,
such as H$\alpha$ and \ion{Ca}{II} 8542 {\AA} lines. Using high spatial
resolution data, it was found that the lifetime of EBs is 2-20 minutes,
and their size can be smaller than 1\arcsec\ (Vissers et al.~\cite{Viss12};
Nelson et al.~\cite{Nel13}; Li et al.~\cite{Li15}). The temperature
increase of EBs around the temperature minimum region (TMR) is about
600-1500 K (Georgoulis et al.~\cite{Geor02}; Fang et al.~\cite{Fang06};
Hong et al.~\cite{Hong14}; Berlicki et al.~\cite{Berl14}). Recently,
using high-resolution H$\alpha$ and \ion{Ca}{II} 8542 {\AA} spectra
obtained by the 1.6m New Solar Telescope (NST), Li et al.~(\cite{Li15})
found that the temperature increase can be about 3000 K even for three
smallest EBs. The energy of EBs is estimated to be in the range of
10$^{25}$ -- 10$^{27}$ ergs (e.g. Georgoulis et al.~\cite{Geor02};
Fang et al.~\cite{Fang06}; Li et al.~\cite{Li15}). It should be
emphasized that there are some ``pseudo-EBs'', being bright points
in some images, but probably manifestations of deep radiation escape
(Rutten et al.~\cite{Rutt13}; Vissers et al. ~\cite{Viss15}).
Thus, the best way to identify EBs is using spectral data.

Recently, with the Interface Region Imaging Spectrograph (IRIS;
De Pontieu et al.~\cite{DeP14}), some authors found that there are small-scale bright
regions observed in the 1400 {\AA} and 1330 {\AA} images, called
IRIS bombs (IBs) (Peter et al.~\cite{Pete14}; Tian et al.~\cite{Tian16}),
which have a local heating in the photosphere up to $\sim$ 8 $\times 10^4$ K
under the assumption of collisional ionization equilibrium (Peter et al.
~\cite{Pete14}) or 1-2 $\times 10^4$ K under the local thermodynamic
equilibrium (LTE) assumption (Rutten et al.~\cite{Rutt16}). Thus,
the relationship between the EBs and IBs becomes an interesting question.
Vissers et al.(~\cite{Viss15}) studied five EBs and found that strong
EBs can produce IB-type spectra. Kim et al.~(\cite{Kim15}) also found
the connection between an IB and an EB. Recently, Tian et al.
(~\cite{Tian16}) used IRIS and the Chinese New Vacuum Solar Telescope
(NVST; Liu et al.~\cite{Liu14}) data, and identified 10 IBs. Among them,
3 are obviously and 3 others are possibly connected to EBs, and the
remaining four IBs are not EBs. They concluded that some EBs connected
to IBs can be heated to 1$\sim$8 $\times 10^4$ K, which is much hotter than
that obtained from non-LTE modeling of EBs. Thus, whether EBs can be
heated to such a high temperature is a hot topic.

In this paper, we use non-LTE modeling to investigate some characteristics
of EB models with different temperatures, and compare them with observations,
which were performed with the largest aperture solar telescope in the world, the
1.6 meter off-axis New Solar Telescope (NST) (Goode et al.~\cite{Goode12};
Cao et al.~\cite{Cao10}) at the Big Bear Solar Observatory (BBSO).
The method of non-LTE modeling is described in \S\, \ref{sect2}. The resulted
characteristics of EB models are given in \S\, \ref{sect3}. General discussion
and conclusion are given in \S\, \ref{sect4}.

\section{Semi-empirical modeling of EB models with different temperatures}\label{sect2}

We use the non-LTE method as described in the paper of Fang et al.(~\cite{Fang06}).
That is, prescripting a semi-empirical temperature distribution of an EB model,
we solve the statistical equilibrium equation, the radiative transfer equation,
the hydrostatic equilibrium, and the particle conservation
equations iteratively. A four-level hydrogen atom and a five-level \ion{Ca}{II}
atom are taken. All the bound-free and the free-free transitions of the hydrogen
atom and negative hydrogen (H$^-$) are included. The converging criteria is that
the relative difference of the mean intensities between the last two iterations
is less than 10$^{-7}$ and 10$^{-9}$ for hydrogen and calcium atoms, respectively.
Then the H$\alpha$ and \ion{Ca}{II} 8542 {\AA} line profiles and continuum
intensity can be calculated.

As the typical cases, we take three EB models with different temperatures
around the TMR. The models labeled EB15000 and EB10000 have a maximum
temperature of 15000K and 10000K, respectively. The third one labeled EB0955
has a maximum temperature of 7000K, which is about the same as that
given by the empirical model labeled No.2 EB in our recent paper
of Li et al.(~\cite{Li15}). The empirical model of EB0955 can well reproduce the
H$\alpha$ and \ion{Ca}{II} 8542 {\AA} line profiles observed by the Fast Imaging Solar
Spectrograph (FISS) (Chae et al.~\cite{Chae13}) of BBSO/NST on 2013 June 6
at 09:55 UT. Compared with the VALC quiet-Sun model (Vernazza et al.~\cite{Ver81}),
the temperature increases for the three EB models are 11000K, 6000K, and 3000K,
respectively.

\begin{figure}
\centerline{\includegraphics[width=0.7\textwidth,angle=90.0]{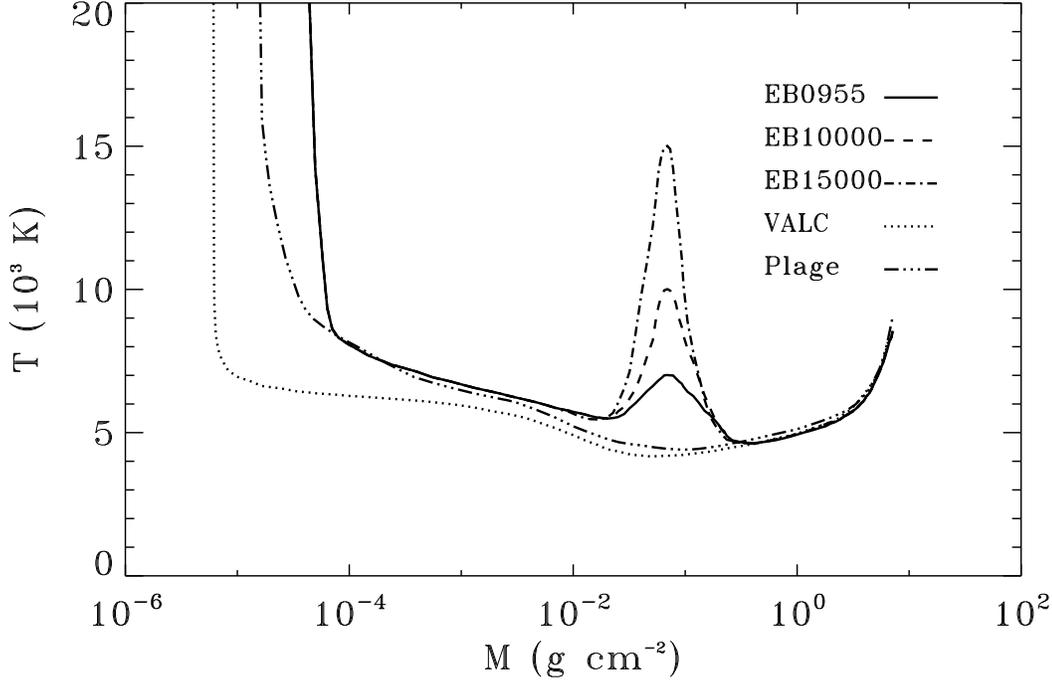}}
 \caption{Temperature distributions in the three EB models:
     EB0955 ({\it solid line}), EB10000 ({\it dashed line}) and EB15000
     ({\it dashed-dotted line}), compared to that of the plage model
     ({\it dashed three-dotted line}) given by Fang et al.(~\cite{Fang01}), and
     that of the quiet-Sun model (i.e., the VALC model,  {\it dotted line})
     given by Vernazza et al.(~\cite{Ver81}).}\label{fig1}
\end{figure}

Figure \ref{fig1} gives the temperature distributions for the three models.
For comparison, we also plot the temperature distributions in the semi-empirical
model for plages (denoted by ``Plage'') given by Fang et al.(~\cite{Fang01})
and for the VALC quiet-Sun model (denoted by ``VALC''). In the figure, M is
the column mass density.

%Figure

\begin{figure}
\centerline{\includegraphics[width=1.\textwidth]{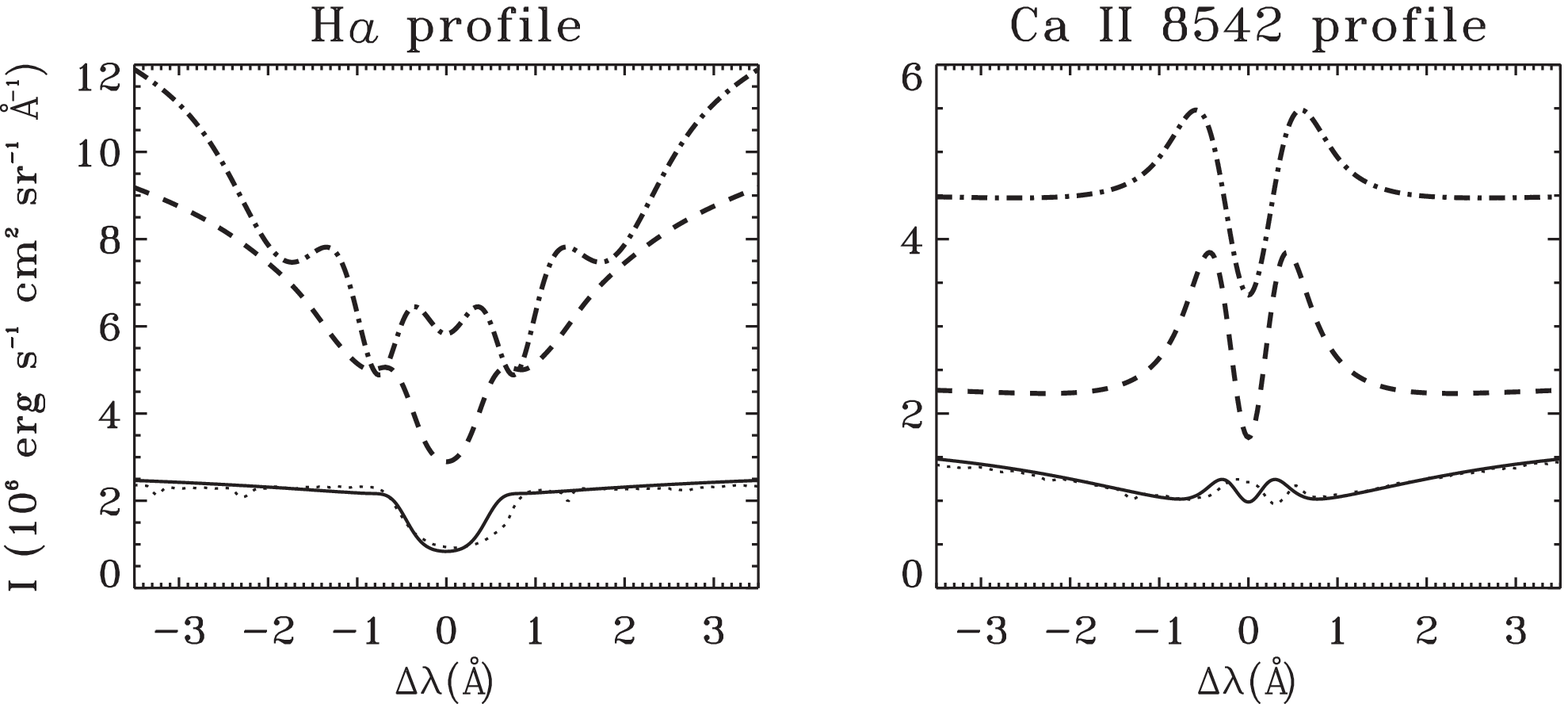}}
 \caption{Comparison between the observed ({\it dotted lines}) and
 computed H$\alpha$ and \ion{Ca}{II} 8542 {\AA} line profiles for the EB0955
 ({\it solid lines}), the EB10000 ({\it dashed lines}) and the EB15000
 ({\it dashed-dotted lines}). All theoretical profiles are convolved
 with a macroturbulence velocity of 8 km $s^{-1}$.}\label{fig2}
\end{figure}

Figure \ref{fig2} gives both the observed and computed H$\alpha$
and \ion{Ca}{II} 8542 {\AA} line profiles for the three EB models. It can be seen that
the computed profiles of the EB0955 can well match the observed ones, while
that of the EB10000 and EB15000 are much stronger and broader than that
given by observations.

\begin{figure}
\centerline{\includegraphics[width=0.8\textwidth]{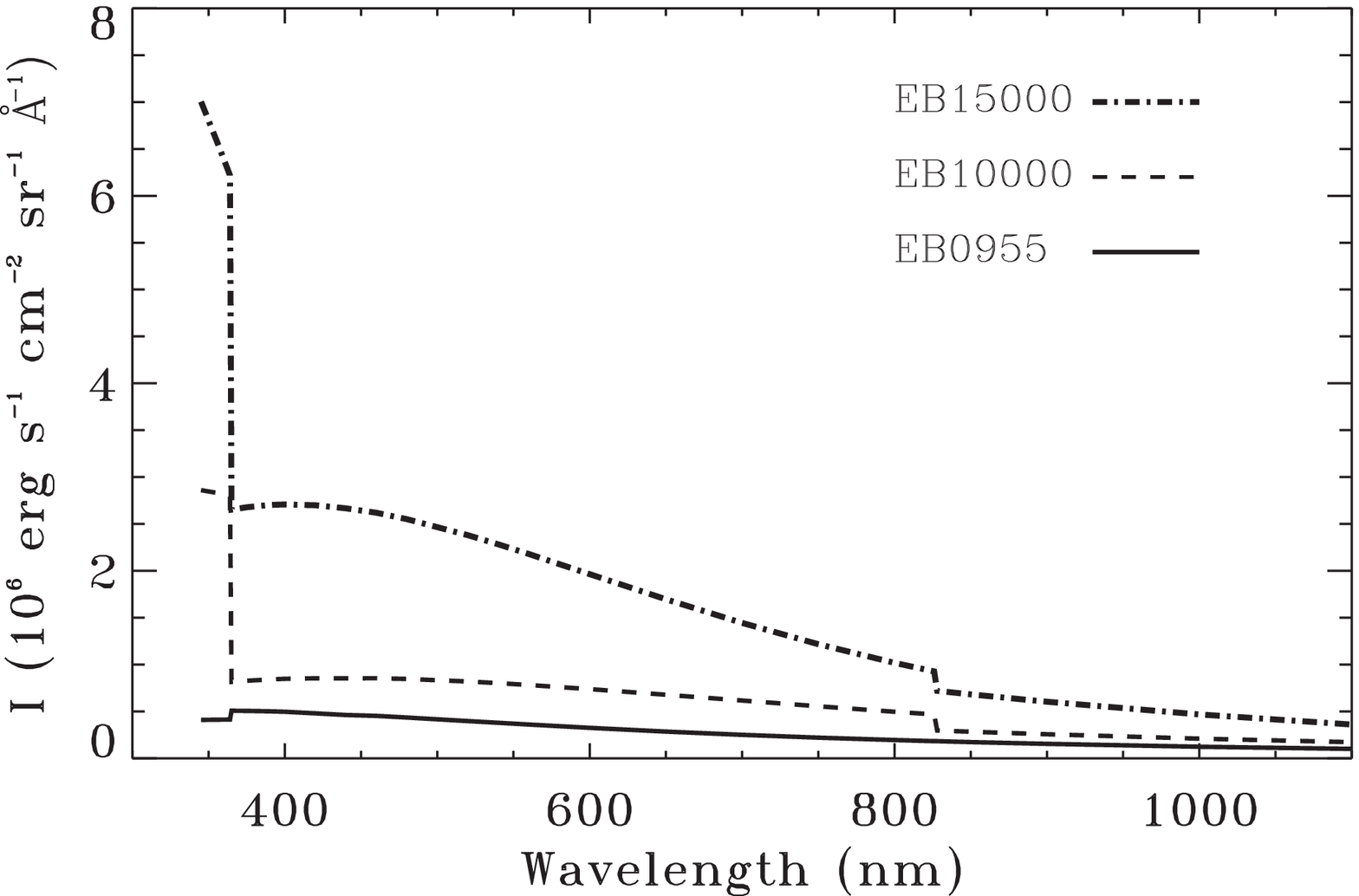}}
 \caption{Continuum emissions for the three EB models.}\label{fig3}
\end{figure}

Using the non-LTE method mentioned above, we can calculate the continuum emission
for the three models. Figure \ref{fig3} gives the result. It can be seen that
the continuum emissions of the EB10000 and EB15000 are much stronger than
that of the EB0955. Particularly, there would be a very strong Balmer jump for both
EB10000 and EB15000. Obviously, such strong continuum emissions and very strong
Balmer jump are never observed in EB's spectra.

\section{Radiative losses}\label{sect3}

We can further estimate the radiative losses of the three EB models by using
the non-LTE modeling. The method is similar to that in Fang et al.(~\cite{Fang06}).
That is, considering that the main heating regions of three EB models are around
the TMR, we can use the following equation to estimate
the total radiative losses $E_r$ of the EB models:

\begin{equation}
E_r = D  A_{EB} {R_{EB}\over 4} \,\,,
\end{equation}

\noindent
where

$R_{EB} = \int_{h_1}^{h_2} {R_r}\, d {h}$ \\

\noindent
is the height-integrated radiative cooling rate per unit area in the main
heating region [$h_1$, $h_2$]. $h_1$ and $h_2$ are the lower and the upper
heights of the heated region, respectively. $R_r$ is the peak rate of
radiative losses in units of erg cm$^{-3}$ s$^{-1}$. Considering that
the rate of radiative losses changes during the EB lifetime $D$, we
take 1/4 of the peak value as the mean rate. $A_{EB}$ is the area
of the EB.  We use the empirical formula given by Jiang et al.(~\cite{Jiang10}),
which is a modification of the formula of Gan et al.(~\cite{Gan90}) and
is more suitable for the small-scale activities. It is given as follows:

\begin{equation}
R_r= n_H n_e ({\alpha_1 (h)}+{\alpha_2 (h)}) f(T)\,,
\end{equation}

\noindent
where

\begin{figure}
\centerline{\includegraphics[width=0.8\textwidth]{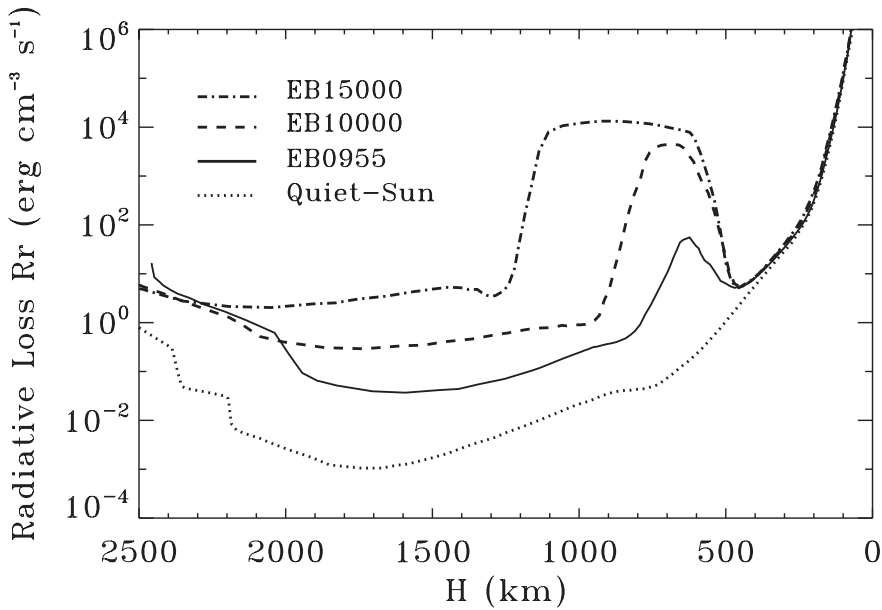}}
 \caption{Peak radiative losses for the three EB models.}\label{fig4}
\end{figure}

$log ~{\alpha_1 (h)} = 1.745 \times 10^{-3} h -4.739$\,,

$\alpha_2 (h) = 8.0 \times 10^{-2} e^{-3.701\times 10^{-2} h}$\,,

$f(T) = 4.533 \times 10^{-23} {(T/10^4)}^{2.874}$\,, \\

\noindent
where $h$ is the height in kilometer. Thus, using the non-LTE calculation results,
we can obtain the radiative losses for the three EB models. Figure \ref{fig4} gives
the result. It is clear that the radiative losses of the EB10000 and EB15000 are
about two or three orders of magnitude higher than that of the EB0955. This is due to
their very high electron density in the lower atmosphere, which is caused by photoionization
produced by high Balmer and Paschen emissions in the EB region, as shown in
Figure \ref{fig3}.

Further, if assuming that the main mechanism of the energy losses is the
radiative one, and taking the e-folding time of the maximum temperature decreasing
as the lifetime of the EBs, then, for a given energy $E$ of an EB,
we can estimate its lifetime by the formula:

\begin{equation}
D = {{4 \times E} \over {A_{EB} R_{EB}}}\, .
\end{equation}

\noindent
From the non-LTE calculation, we can obtain all necessary quantities
for estimating the durations of the three EB models. For comparison, we
assume that all the EBs have the typical size of EBs being
$0 \farcs 5 \times 0 \farcs 5$. We take $h_1$ = 337 km, and
$h_2$ = 2044 km for all three EBs. Considering the rapid decrease
of the hydrogen density with height, we neglect the contribution
from the higher layers.

We take the total radiative energy $E$ = 1 $\times 10^{26}$ ergs,
which is a typical value for an EB (e.g. Fang et al.~{\cite{Fang06};
Li et al.~\cite{Li15}). Using Eq. (3), we can estimate the lifetimes of
the three EBs, which are listed in Table \ref{tab1}. It can be
seen that the lifetimes of EB10000 and EB15000 would be two or three
orders of magnitude shorter than the EB0955, due to their very high
radiative losses, and are not comparable with observations. Of course,
for some very bright EBs the total radiative energy could be higher,
say, 1 $\times 10^{27}$ ergs. In this case, the lifetimes of the
high-temperature EBs are still very short. Moreover, for an EB having
high temperature, the energy losses by heat conduction should be
considered. It will decrease the EB lifetime further.

\begin{table*}
\begin{center}
\caption{Characteristics for the three EB models}\label{tab1}
\end{center}
\begin{center}
\begin{tabular}{c c c c c}
\hline
 EB & Assuming size ($x \times y$) & $ T_{max}$ & $R_{EB}$ & Lifetime \\
    & (arcsec) & (K) & (erg cm$^{-2}$ s$^{-1}$)  & (s) \\
\hline
EB0955  & 0.5$\times$ 0.5  &  7000 & 8.21$\times 10^{8}$   & 371  \\
EB10000 & 0.5$\times$ 0.5  & 10000 & 7.26$\times 10^{10}$  & 4.19 \\
EB15000 & 0.5$\times$ 0.5  & 15000 & 6.08$\times 10^{11}$  & 0.50 \\
\hline
\end{tabular}
\end{center}
\end{table*}
%============================================================================

\section{Discussion and Conclusion}\label{sect4}

Using non-LTE theory, we calculate the H$\alpha$ and \ion{Ca}{II} 8542 {\AA}
line profiles, as well as the continuum emissions, for three EB models with
different temperatures around the TMR. Our results
clearly reveal that the EB models with the maximum temperature over 10000K
will produce not only much stronger H$\alpha$ and \ion{Ca}{II} 8542 {\AA} line
profiles, but also very strong continuum emissions, particularly in the Balmer
continuum. These line profiles and continuum emissions are not comparable
with an EB model denoted as EB0955, which reproduces well the observations
by BBSO/NST on 2013 June 6 at 09:55 UT. It should be emphasized that all
these characteristics produced from EB10000 and EB15000 are
never reported in EB spectral observations. Moreover, as mentioned above,
we regard the e-folding time of the EBs as its lifetime, and the calculation
is approximate since the radiative cooling rate changes non-linearly with
temperature. Nevertheless, our results at least indicate that an EB with
a temperature increase over 10000 K around the TMR, if actually produced,
cannot persist enough long with its high-temperature character.
This is certainly not matching the EB observations.
In a word, it is unlikely that the maximum temperature
of EBs around the TMR could excess 10000K.  We notice that recently
Reid et al.~(\cite{Reid17}) used RADYN 1-dimensional radiative transfer code and got
the conclusion that the presence of superheated regions in the photosphere
($>$10,000 K) is not a plausible explanation for the production of EB signatures.
It is consistent with our result.

It is thought that magnetic reconnection in the photosphere
or lower chromosphere could be a mechanism for EBs (Ding et al.
~\cite{Ding98}; Georgoulis et al.~\cite{Geor02}; Fang et al.
~\cite{Fang06}; Pariat et al.~\cite{Par07}; Isobe et al.~\cite{Iso07};
Watanabe et al.~\cite{Wata11}; Yang et al.~\cite{Yang13};
Nelson et al. ~\cite{Nel13}). We have performed two-dimensional
numerical MHD simulations on the magnetic reconnection in the
solar lower atmosphere (Chen et al. ~\cite{Chen01}; Jiang et al.
~\cite{Jiang10}; Xu et al.~\cite{Xu11}).
Our results indicated that magnetic reconnection in the lower solar
atmosphere can explain the temperature enhancement of about 600-3000 K and
the lifetime of EBs. Recently, Ni et al. (~\cite{Ni16}) performed 2.5
dimensional MHD simulation and indicated that both the high temperature
($\geq$ 8 $\times 10^4$ K, if the plasma $\beta$ is low) and low temperature
($\sim 10^4$ K, if $\beta$ is high) events can happen around the TMR.
However, they did not include the non-equilibrium ionization effect,
which is certainly important for the high temperature case (Chen et al. ~\cite{Chen01}).
Nevertheless, their result implies that the high and low temperature
events can not happen at the same place. It is true that IBs need
a high temperature of over 10000 K to explain the enhanced
EUV lines. However, according to our result, the maximum temperature
of EBs around the TMR is unlikely higher than 10000 K. This implies
that IBs and EBs are unlikely to happen at exactly the same place.
There are indeed some IBs that are closely connected to EBs in
observations (e.g., Tian et al.~\cite{Tian16}. They may reflect different
temperature components that are produced by a common magnetic
reconnection process. In fact, previous numerical simulations have
revealed that both low and high temperature components can exist
in the case of emerging flux reconnecting with pre-existing magnetic
field (Yokoyama et al.~\cite{Yok95}; Jiang et al.~\cite{Jiang12}).
Another possibility is that the jets produced by some EBs (see e.g.
Nelson et al.~\cite{Nel15}) could heat the upper atmosphere via waves
or shocks. In this case, the high temperature region would be higher
than the low temperature one. Then the apparent coincidence of both
phenomena could only be due to the project effect. For checking
this point, a coordinated observation between IRIS and a large-aperture
ground-based telescope in a region close to the limb may be useful.
Anyway, for a clearer explanation of the relationship between EBs and IBs,
we need more observations with higher spatial and temporal resolutions
in the future.

Based on the analysis of the three EB models, we draw the conclusion
as follows:

According to our results of non-LTE calculation and semi-empirical modeling,
it is unlikely that the maximum temperature of EBs around the TMR
could exceed 10000K, otherwise the resulted H$\alpha$ and \ion{Ca}{II} 8542 {\AA}
line profiles, as well as the continuum emissions, would be too strong to be
comparable with the EB observations. Particularly, due to very high
radiative losses in the high temperature models (EB10000 and EB15000), the lifetime
of these events would be very short, which is inconsistent with any EB observations.

\begin{acknowledgements}
We thank a lot to the anonymous Referee for his/her valuable suggestions.
This work is supported by the National Natural Science Foundation of
China (NSFC) under the grants 11533005, 11025314, 13001003, 11203014,
11103075 as well as NKBRSF under grants 2014CB744203.
\end{acknowledgements}

\end{document}